\documentstyle[12pt]{article}
\newcommand{\bea}{\begin{eqnarray}}
\newcommand{\eea}{\end{eqnarray}}
\newcommand{\beq}{\begin{equation}}
\newcommand{\eeq}{\end{equation}}
\newcommand{\bay}{\begin{array}}
\newcommand{\eay}{\end{array}}

\def \cn{Collaboration}
\def \ite{{\it et al.}}
\begin{document}
\rightline{CERN-TH/2003-033}
\rightline{hep-ph/03mmnnn}
\rightline{February 2003}
\bigskip
\bigskip
\centerline{\bf P AND CP VIOLATION IN $B$ PHYSICS}
\bigskip
\centerline{\it Michael Gronau\footnote{Permanent Address: Physics Department,
Technion -- Israel Institute of Technology, 32000 Haifa, Israel.}} 
\centerline{\it Theory Division, CERN} 
\centerline{\it CH--1211, Geneva 23, Switzerland}
\bigskip
\centerline{\bf ABSTRACT}
\vskip 2cm  

\begin{quote}
While the Kobayashi--Maskawa single phase origin of CP violation passed its first crucial
precision test in $B\to J/\psi K_S$, the chirality of weak $b$-quark couplings has not 
yet been carefully tested. We discuss recent proposals for studying the chiral and 
CP-violating structure of these couplings in radiative and in hadronic $B$ decays.
\end{quote}

\vskip 5.0cm
\centerline{\it Invited talk at the IXth International Symposium on Particles, Strings
and Cosmology}
\centerline{\it Tata Institute of Fundamental Research Colaba, Mumbai, Jan. 3$-$8, 2003}
\newpage

\section{Introduction}
It has often been stated that CP violation is among the least tested and poorly 
understood properties of the Standard Model (SM) of electroweak and strong 
interactions. CP asymmetries measured in meson decays are accounted for by 
arbitrary complex Yukawa couplings, yielding a single phase in the 
Cabibbo--Kobayashi--Maskawa (CKM) matrix 
\cite{Cab,KM}. The Kobayashi--Maskawa (KM) model for CP violation was suggested 
thirty years ago \cite{KM} to explain CP non-conservation in $K$ decays. Recently it 
passed in a remarkable way its first crucial \cite{sanda} precise \cite{gold} test in 
$B$ decays when a large CP asymmetry was measured in $B^0 \to J/\psi K_S$ 
\cite{psiKS}. This opens a new era, in which the model is expected to be scrutinized through 
a variety of other $B$ and $B_s$ decay asymmetries. Impressive progress has already 
been made in search for asymmetries in several hadronic $B$ decays, including $B^0(t) \to 
\pi^+\pi^-$ \cite{pipi}, $B^{0,\pm} \to K \pi$ and $B^{\pm} \to DK^{\pm}$ 
\cite{DK}. Current measurements are approaching the level of tightening bounds on the 
CP-violating phase $\gamma$ \cite{MGR}. These and forthcoming measurements of $B_s$ decays 
\cite{Bs} will enable a cross-check of the KM model. Two complementary tools for 
studying hadronic decays, an isospin \cite{isospin} and flavor SU(3)-based approach 
\cite{GHLR} and a QCD-factorization approach \cite{BBNS}, can be checked and refined when 
confronted by data. This is expected to establish the KM hypothesis at a high precision, 
up to a point where deviations will hopefully be observed from this simple picture. 

The CP transformation consists of a product of parity (P) and charge-conjugation
(C). In the SM, parity is violated in a maximal way by assuming
that left-handed chiral fermion fields transform as doublets under the SU(2) gauge
group, while right-handed fermions transform as singlets. This left--right asymmetry, 
which is introduced by hand in order to account for the observed low energy 
phenomenology, may ultimately disappear at high energies in a left--right symmetric 
theory  \cite{LR}. This theory would hopefully explain the origin and pattern of fermion 
masses, mixing and CP-violating phases, which in the current model are represented by 
arbitrary complex Yukawa couplings. In such a theory, parity and CP violation would 
have a common origin in a new type of interaction responsible for fermion masses. If 
parity violation and the quark mass hierarchy have a common origin, then it may be
anticipated that right-handed couplings grow with quark masses and are 
larger for $b$ quarks than for $s$ quarks, in contrast to the measured left-handed weak 
couplings. In this case the very small $b$ couplings \cite{PDG}, 
$|V_{cb}| = 0.04,~|V_{ub}| = 0.003$, would be sensitive to right-handed interactions. 
In order for right-handed $b$ couplings to be observable in the near future, the left--right 
symmetry scale would have to be around a TeV or several TeV \cite{RMAP}, not far above the 
electroweak scale. In the absence of good experimental tests for the chirality of $b$ 
quark couplings \cite{MG}, such a scenario cannot be ruled out and adequate tests are 
desirable. This report will focus on several such tests. 

\medskip
I will discuss photonic and hadronic $B$ decays, the study of which serves two purposes:
\begin{enumerate}
\item Examining the chiral structure of $b$-quark couplings in P-odd observables.
\item Testing CKM phases in CP-violating observables. 
\end{enumerate}  
Since the second topic has recently been the subject of frequent discussions \cite{CP}, 
I will pay more attention to the first aspect, discussing it in Sections 2 and 3. The second 
topic will be addressed in Section 4 while Section 5 will conclude.

\section{The photon polarization in radiative $B$ decays}
In radiative $b$ quark decays, $b \to s \gamma$, the $s$ quark, which couples to a 
$W$ in the loop, is left-chiral. Therefore, the photon is predominantly left-handed.
The effective Hamiltonian for $b \to s\gamma$ contains a dominant dipole-type
operator given by 
\cite{hurth}
\bea
{\cal H}_{\rm rad} &=& -\frac{4G_F}{\sqrt2} V_{tb} V_{ts}^*\left( C_{7L}{\cal
O}_{7L} +  C_{7R}{\cal O}_{7R}\right)~,\\
{\cal O}_{7L,R} &=& \frac{e}{16\pi^2} m_b
\bar s\sigma_{\mu\nu}\frac{1 \pm \gamma_5}{2} bF^{\mu\nu}~~,
\eea
where the Wilson coefficients $C_{7L}$ and $C_{7R}$ describe the amplitudes  
for left- and right-handed photons, respectively. 
In the SM one has $C_{7R}/C_{7L} =  m_s/m_b$. Defining the photon 
polarization in $b \to s\gamma$,
\beq
\lambda_\gamma \equiv  \frac{|C_{7R}|^2 - |C_{7L}|^2}{|C_{7R}|^2 + |C_{7L}|^2}~~, 
\eeq
the SM predicts 
\beq\label{lambda}
\lambda_\gamma = -1~,
\eeq
where ${\cal O}(m^2_s/m^2_b)$ corrections are expected to be at the per cent level. 
Also four-quark operators, $O_{1,2}$ with (V--A)(V--A) structure, contribute to a 
dominantly left-handed photon, whereas contributions from penguin operators, 
$O_{3-6}$, with a different chiral structure, involve much smaller Wilson coefficients 
and can be safely neglected.   

As we will now argue \cite{GP}, the prediction (\ref{lambda}) for the photon 
polarization holds also in the presence of hadronic effects in exclusive radiative decays 
in which the final hadronic system carries well-defined spin and parity. 
Let us consider the decay of a $\bar B$ meson into a strangeness $+1$ state $X_s$ 
with spin-parity $J^P$, and let us denote hadronic and photonic states with 
helicities $\pm 1$ by $X^{R,L}_s$ and $\gamma_{R,L}$, respectively. One 
clearly has
\beq
\langle X^L_s \gamma_L|{\cal O}_{7R}|\bar B\rangle =
\langle X^R_s \gamma_R|{\cal O}_{7L}|\bar B\rangle = 0~~,
\eeq
while parity and rotational invariance of the strong interactions imply
\beq
\langle X^R_s \gamma_R|{\cal O}_{7R}|\bar B\rangle = (-1)^{J-1}P 
\langle X^L_s \gamma_L|{\cal O}_{7L}|\bar B\rangle~~.
\eeq
Defining weak radiative amplitudes for left- and right-polarized photons, 
\beq
c_i \equiv \langle X^i_s \gamma_i|{\cal H}_{\rm rad}|\bar B\rangle~~,~~~~i = L,R~~,
\eeq
one obtains
\beq
\frac{c_L}{c_R} = (-1)^{J-1}P \frac{C_{7L}}{C_{7R}}~~\Rightarrow~~
\frac{|c_R|^2-|c_L|^2}{|c_R|^2+|c_L|^2} = \frac{|C_{7R}|^2 - |C_{7L}|^2}
{|C_{7R}|^2 + |C_{7L}|^2}~~.
\eeq
Thus, both in inclusive radiative $B$ decays and in exclusive decays to $J^P$ states, 
there is a prediction for the photon polarization in terms of ratio of Wilson coefficients. 
The prediction (\ref{lambda}) holds in the SM at the per cent level. It is amusing to note 
that, while the current calculation of the measured inclusive rate has not yet 
achieved a level of $10\%$ in spite of great efforts \cite{hurth}, the more precisely 
predicted photon chirality has not yet been put to an experimental test.

The photon polarization prediction is very sensitive to new physics effects.
In several extensions of the SM the photon in $b\to s\gamma$ 
acquires an appreciable right-handed component due to chirality flip
along a heavy fermion line in the electroweak loop. Two well-known 
examples of such extensions are the left--right-symmetric model and the 
unconstrained Minimal Supersymmetric Standard Model. In the first model 
chirality flip along the $t$-quark line in the loop involves $W_L$--$W_R$ 
mixing \cite{LRM}, while in the second scheme
a chirality flip along the gluino line in the loop involves left--right squark 
mixing \cite{SUSY}. In both types of models it was found that, in certain 
allowed regions of the parameter space, the photons emitted in $b\to s\gamma$ 
can be largely right polarized without affecting substantially
the SM prediction for the inclusive radiative decay rate. This situation 
calls for an independent measurement of the photon polarization.

Several ways were proposed for measuring photon helicity effects in
radiative $B$ decays. We will briefly describe three early suggestions. 
The first two proposals are sensitive to interference between left and right 
polarization amplitudes; they can be used to forbid a large interference 
at present $B$ factories. This may exclude certain parameters in 
left--right-symmetric and SUSY models. The third method, which measures the 
photon polarization directly, requires an extremely high luminosity $e^+ e^-$ $Z$ factory.
Finally, we will focus our attention on a recent proposal to measure the photon 
polarization, which is feasible at currently operating $B$ factories.

\subsection{CP asymmetry in $B^0(t) \to X^{CP}_{s(d)}\gamma$} 
Consider the time-dependent rate of \cite{AGS} $B^0(t) \to X^{CP}_{s(d)}\gamma$, where 
$X^{CP}_s = K^{*0} \to K_S\pi^0$ or $X^{CP}_d = \rho^{0} \to \pi^+\pi^-$. The 
time-dependent CP asymmetry follows from an interference between $B^0$ and $\bar B^0$ 
decay amplitudes into a common state of definite photon polarization, and is 
proportional to $C_{7R}/C_{7L}$. For instance, in the SM the asymmetry in $B^0(t) 
\to f,~ f =K^{*0}\gamma\to (K_S\pi^0)\gamma$, is given by
\beq
{\cal A}(t) \equiv \frac{\Gamma(B^0(t) \to f) - \Gamma(\bar B^0(t) \to f)}
{\Gamma(B^0(t) \to f) + \Gamma(\bar B^0(t) \to f)}
= {2C_{7L} C_{7R} \over C_{7L}^2 + C_{7R}^2}\sin 2\beta\sin(\Delta mt)~.
\eeq
The ratio $C_{7R}/C_{7L}$ is expected to be a few per cent in the SM,
whereas it may be much larger in extensions of the SM \cite{AGS}.

\subsection{Angular distribution in $\bar B\to \bar K^*\gamma \to \bar K\pi e^+ e^-$}
Consider the decay distribution in this process as a function of the angle $\phi$ 
between the $\bar K\pi$ and $e^+e^-$ planes, where the photon can be virtual 
\cite{Kim} or real, converting in the beam pipe to an electron--positron pair 
\cite{GrPi}. The $e^+e^-$ plane acts as a polarizer, the distribution in $\phi$
is isotropic for purely circular polarization, and the angular distribution is 
sensitive to interference between left and right polarization. One finds
\beq
\frac{d\sigma}{d\phi} \propto 1 + \xi {C_{7L} C_{7R} 
\over C_{7L}^2 + C_{7R}^2}\cos(2\phi + \delta)~,
\eeq
where the parameters $\xi$ and $\delta$ are calculable and involve hadronic physics.
 
\subsection{Forward--backward asymmetry in $\Lambda_b \to \Lambda\gamma \to p\pi\gamma$}
The forward--backward asymmetry of the proton with respect to the $\Lambda_b$ in the 
$\Lambda$ rest frame is proportional to the photon polarization 
\cite{MaRe}. Using polarized $\Lambda_b$'s from extremely high luminosity 
$e^+e^-$ $Z$ factories, one can also measure the forward--backward asymmetry of the
$\Lambda$ momentum with respect to the $\Lambda_b$ boost axis \cite{HK}. This 
asymmetry is proportional to the product of the $\Lambda_b$ and photon polarizations.

\subsection{Angular distribution in $B\to K_1(1400)\gamma \to K\pi\pi\gamma$}
Radiative decays into an excited tensor meson resonance state, $K^*_2(1430)$, 
identified through its $K\pi$ decay mode, were observed by the CLEO \cite{CLEOK**} 
and Belle \cite{Belle} collaborations. The Belle collaboration also observed a
$K\pi\pi$ resonant structure above an invariant mass of 1.2 GeV, consistent with a 
large mixture of $K_1(1400) \to K^*\pi$. We will show that a detailed angular 
analysis of $K\pi\pi\gamma$ events can be used to study the photon polarization.

First, we point out \cite{GGPR} that, in order to measure the photon polarization 
$\lambda_\gamma$ in radiative $B$ decays through the recoil hadron distribution, 
one requires that the hadrons consist of at least three particles. This necessary
condition follows from the simple fact that $\lambda_\gamma$ is parity-odd. Consequently, 
the hadronic quantity multiplying $\lambda_\gamma$ in the decay distribution must be P-odd.
The pseudoscalar quantity, which contains the smallest number of hadron momenta, is 
a triple product that requires three hadrons recoiling against the photon.
The idea is then to measure an expectation value $\langle \vec 
p_{\gamma}\cdot (\vec p_1\times \vec p_2)\rangle$, where $\vec p_1$ and $\vec p_2$
are momenta of two of the hadrons in the centre-of-mass frame of the recoil hadrons. 
Since the triple product is also time-reversal-odd, a non-zero expectation value requires a phase 
due to final state interactions. The final state interaction phase is calculable in the 
special case that the decay occurs through two isospin-related intermediate $K^*$ 
resonance states. This is the case that we will discuss now \cite{GP,GGPR}.

Consider the decays $B^+ \to K^+_1(1400)\gamma$ and $B^0 \to K^0_1(1400)\gamma$, 
where $K^+_1$ and $K^0_1$ are observed through 
\bea
K^+_1(1400)\to \left\{
\begin{array}{c}
 K^{*+}\pi^0 \\
 K^{*0} \pi^+ 
\end{array}
\right\} \to K^0 \pi^+ \pi^0,~
K^0_1(1400)\to \left\{
\begin{array}{c}
 K^{*+}\pi^- \\
 K^{*0} \pi^0 
\end{array}
\right\} \to K^+ \pi^- \pi^0~,
\eea
with a branching ratio ${\cal B}(K_1\to K^*\pi) = 0.94\pm 0.06$ \cite{PDG}.
In both charged and neutral $K_1$ decays, two Breit--Wigner amplitudes interfere 
through 
intermediate $K^{*+}$ and $K^{*0}$. Decays to $\rho K$, with ${\cal B}(K_1\to \rho K) 
= 0.03\pm 0.03$ \cite{PDG}, will be neglected at this point. The two $K^*$ 
amplitudes are related by isospin; therefore phases other than
the two Breit--Wigner phases cancel. The decay $K_1 \to K^*\pi$ is dominated 
by an $S$ wave and involves a small $D$ wave amplitude, where the $D/S$ ratio of rates is
$|A_D/A_S|^2 = 0.04 \pm 0.01$ \cite{PDG}. Using Lorentz invariance, it is straightforward 
to write down the decay amplitude for $B\to (K\pi\pi)_{K_1}\gamma$, and to calculate
the decay distribution in the $K_1$ rest frame \cite{GP,GGPR},
\beq
\frac{d\Gamma}{ds_{13}ds_{23}d\cos\theta} \propto |\vec J|^2(1 + \cos^2\theta)
+ \lambda_{\gamma} 2{\rm Im}\left (\hat n\cdot (\vec J\times\vec J^*)\right )
\cos\theta~.
\eeq
The vector $\vec J$ is an antisymmetric function of the two pion momenta 
$p_1,~p_2$ and of the $K$ momentum $p_3$. It involves a Breit--Wigner function
$B(s) = \left( s - m_{K^*}^2 - im_{K^*} \Gamma_{K^*}\right)^{-1}$ of $s_{i3}=(p_i 
+ p_3)^2~(i=1,2)$. The angle $\theta$ lies between the 
normal to the decay plane $\hat n\equiv (\vec p_1 \times \vec p_2)/|\vec p_1 \times 
\vec p_2|$ and $-\vec p_\gamma$, all measured in the $K_1$ rest frame. A useful
definition of the normal is in terms of the slow and fast pion momenta, $(\vec p_{\rm 
slow}\times \vec p_{\rm fast})/|\vec p_{\rm slow}\times \vec p_{\rm fast}|$. The 
angle between this normal and $-\vec p_\gamma$ will be denoted by $\tilde\theta$.

The decay distribution exhibits an up--down asymmetry of the photon momentum with 
respect to the $K_1$ decay plane. The asymmetry is proportional to the
photon polarization. When integrating over the entire Dalitz plot one finds 
\beq\label{asym}
{\cal A}_{\rm up-down} \equiv \frac{\int_0^{\pi/2}\frac{d\Gamma}{d\cos\tilde\theta}
d\cos\tilde\theta -  \int_{\pi/2}^\pi \frac{d\Gamma}{d\cos\tilde\theta}d\cos\tilde\theta}
{\int_0^\pi \frac{d\Gamma}{d\cos\tilde\theta}d\cos\tilde\theta} = (0.33 \pm 0.05)
\lambda_\gamma~.
\eeq
The uncertainty follows from experimental errors in the $\rho K$ amplitude and in the 
$K^*\pi$ $D$-wave amplitude.
In the SM, where $\lambda_\gamma \approx -1$, the asymmetry is $(33 \pm 5)\%$ and the 
polarization signature is unambiguous: {\em in $B^-$ and $\bar B^0$ decays the photon 
prefers to be emitted in the hemisphere of 
$\vec p_{\rm slow}\times \vec p_{\rm fast}$, while in $B^+$ and $B^0$ it is more 
likely to be emitted in the opposite hemisphere}.

The $K\pi\pi$ invariant mass region above 1.2 GeV contains in addition to $K_1(1400)$ 
other $K$ resonances: an axial-vector $K_1(1270)$, a vector $K^*(1410)$, a tensor 
$K^*_2(1430)$ and higher resonance states. In order to avoid interference between 
$K_1(1400)$ and $K_1(1270)$, one would have to study the region $m(K\pi\pi) = 
1400$--$1550$ 
MeV. The $K^*(1410)$ state leads to no up--down asymmetry, while the $K^*_2(1430)$ 
involves a small asymmetry \cite{GGPR}. These two resonances dilute the overall 
up--down asymmetry in this mass range compared to the asymmetry from $K_1(1400)$; 
however, they do not affect the sign of the asymmetry. 
Thus, a first crude measurement of merely the sign 
of the photon polarization can be made using an integrated sample of $K\pi\pi\gamma$ events. 
In order to improve the efficiency, one would have to isolate events from $K_1(1400)$,
where the asymmetry is largest. A procedure for achieving this goal by studying decay 
angular distributions is described in \cite{GP}. 

Assuming then that $B\to K_1(1400)\gamma \to K\pi\pi\gamma$ events involving one neutral 
pion \cite{charged} can be effectively isolated, one may check the feasibility of measuring 
the photon polarization at currently operating $B$ factories.
A $3\sigma$ measurement of a $33\%$ up--down asymmetry requires about 80 
reconstructed $B\to K_1(1400)\gamma \to K\pi\pi\gamma$ events, including 
charged and neutral $B$ and $\bar B$ decays. Fewer events may be needed using the 
full $\tilde\theta$ dependence. Assuming ${\cal B}(B\to
K_1\gamma) = 0.7\times 10^{-5}$ \cite{BR}, including known $K_1$ and $K^*$ 
branching ratios to the relevant charge states, and allowing another order of magnitude 
for experimental efficiencies, resolution and background, one finds that this number of
reconstructed events can be obtained from a total of $2\times 10^8$ $B\bar B$ pairs, 
including charged and neutrals. This number has already been produced at $e^+e^-$
colliders. In order to measure $\lambda_\gamma$ to about $15\%$, which is the level 
of precision of the theoretical result (\ref{asym}), one would need about 
$10^9~B\bar B$ pairs.

\section{Chirality tests in hadronic $B$ decays}
As mentioned in the introduction, the chirality of weak $b$ quark couplings has 
not yet been put to a test in hadronic $B$ decays. In semileptonic
decays $B\to (D^*\to D\pi) e\bar\nu$, angular decay distributions are 
sensitive to a $W_L$--$W_R$ mixing amplitude, in which the $b\to c$ coupling is $V+A$ 
and the leptonic current is $V-A$. These measurements \cite{CLEOV-A} set upper bounds 
on $W_L$--$W_R$ mixing; however, it cannot distinguish between $W_L$ and $W_R$ exchange, in 
which quark and lepton currents have equal chiralities \cite{GW2}. Such a distinction 
requires a parity-odd observable, which cannot be constructed from two-body hadronic 
$D^*$ decays. 

In the present section we will discuss a method for testing $V-A$ in $b$-quark couplings 
by studying hadronic $B$ decays involving a vector meson and an axial-vector meson,
$B \to D^*a_1$, in which the axial vector meson decays to three pseudoscalars. The idea
is similar to that used to measure the photon or the $K_1$ chirality in 
$B \to K_1\gamma$. In order to calculate helicity amplitudes in $B \to D^*a_1$, we will 
assume factorization and heavy quark symmetry. Predictions following from this assumption 
have recently been tested experimentally in $B$ decays to two vector 
mesons, $B \to D^*\rho$. Let us therefore start with a discussion of this process.

\subsection{Helicity amplitudes in $\bar B^0 \to D^{*+}\rho^-$}
The decays $\bar B^0 \to D^{*+}~(\to D^0\pi^+)~\rho^-~(\to \pi^-\pi^0)$, in which each of the 
two vector mesons decays to two spinless particles whose momenta are measured, can be 
used to study the vector meson polarization \cite{KG}. Using an angular 
momentum decomposition, the decay amplitude can be written in terms of three helicity 
amplitudes, $H_0,~H_+,~H_-$, corresponding to the three polarization states of the vector 
mesons,
\beq\label{A}
A = \frac{3}{2\sqrt {2\pi}}\left [H_0\cos\theta_1\cos\theta_2 + 
\frac{1}{2}(H_+ e^{i\phi} + H_- e^{-i\phi})\sin\theta_1\sin\theta_2\right ]~~.
\eeq
Here $\theta_1$ and $\theta_2$ are the angles between each of the two vector mesons momenta in the 
$B$ rest frame and the momenta of the corresponding daughter particles in the decaying vector 
mesons rest frame; $\phi$ is the angle between the $D^*$ and $\rho$ decay planes.
In the above we use a convention in which the normalized decay angular distribution is given 
by $|A|^2$,
\beq
\frac{1}{\Gamma}\frac{d^3\Gamma}{d\cos\theta_1\cos\theta_2 d\phi} = |A|^2 ~~\Rightarrow~~ 
|H_0|^2 + |H_+|^2 + |H_-|^2 = 1~~.
\eeq
The decay distribution is symmetric under $(H_0,~H_+,~H_-) \to (H^*_0,~H^*_-,~H^*_+)$,
implying that rates into left and right polarizations, $|H_-|^2$ and $H_+|^2$ respectively,
are indistinguishable. Namely, one cannot distinguish in this process between left- and 
right-polarized vector mesons. As mentioned before, this follows from the lack of a 
parity-odd observable when each of the vector mesons decays into two spinless particles. 
Thus, while the rate into longitudinally polarized state $|H_0|^2$ can be  measured, 
only the magnitude of $|H_+|^2 - |H_-|^2$ is measurable, but not its sign.

The following values were reported very recently by the CLEO collaboration \cite{CLEO}:    
\bea\label{CLEO}
|H_0| & = & 0.941 \pm 0.009 \pm 0.006~~,\nonumber \\
|H_+|~{\rm or}~|H_-| & = & 0.107 \pm 0.031 \pm 0.011~~,\nonumber \\
|H_-|~{\rm or}~|H_+| & = & 0.322 \pm 0.025 \pm 0.016~~.
\eea
In their report the collaboration quotes a value for $|H_+|$, which is smaller than $|H_-|$, 
assuming that the $D^*$ predominantly carries the chirality of the $c$ quark, as would 
follow from a $\bar c\gamma_\mu(1 - \gamma_5)b$ coupling. It is important to check this
assumption experimentally.

The three helicity amplitudes $H_{0,\pm}$ can be calculated using 
factorization and heavy quark symmetry \cite{BBNS2}. 
In this approximation, the three normalized amplitudes can be written in terms of meson 
masse. For a $\bar c \gamma_\mu (1 - \gamma_5) b$ current one finds \cite{factor}
\beq\label{H}
H_0 = \left (1 + \frac{4y}{y + 1}\epsilon^2\right )^{-\frac{1}{2}}~~,~~~~~
H_\pm = \left ( 1 \mp \sqrt{\frac{y-1}{y+1}}\right )\epsilon 
\left (1 + \frac{4y}{y + 1}\epsilon^2\right )^{-\frac{1}{2}}~~,
\eeq
where $y \equiv (m_B^2+m_{D^*}^2-m_{\rho}^2)/2m_B m_{D^*} = 1.476,~\epsilon \equiv 
m_\rho/(m_B - m_{D^*}) = 0.236$. Thus, the values
\beq\label{Hi}
|H_0| = 0.940~,~~~~ |H_+| = 0.125~,~~~~|H_-| = 0.318~~,
\eeq
are obtained, in good agreement with (\ref{CLEO}). One expects deviations from
factorization in the amplitudes $H_{\pm}$ which are subleading in $1/m_b$.
The above predictions of the Standard Model apply to $\bar B^0$ decays, while  
in $B^0$ decays the values of $|H_+|^2$ and $|H_-|^2$ are interchanged. 
In the case of a $\bar c \gamma_\mu (1 + \gamma_5) b$ current, the roles of 
$|H_+|$ and $|H_-|$ are interchanged. Thus, while the measured value of $|H_0|$ is in 
excellent agreement with the SM prediction using factorization and heavy quark symmetry,  
measurements of $|H_{\pm}|$ cannot distinguish between $V-A$ and $V+A$ currents. 
The present experimental precision allows also for 
an admixture of the two chiralities in the $b \to c$ coupling. 

\subsection{Chirality test in $\bar B^0 \to D^{*+}a^-_1$}
A large sample of $18000 \pm 1200$ partially reconstructed $\bar B^0 \to D^{*+}a^-_1$ events, 
combining this mode with its charge-conjugate, was reported very recently by the BABAR 
collaboration \cite{BABARa1}. The $a_1$ was reconstructed via the decay chain $a^-_1 \to 
\rho^0\pi^-,~\rho^0\to \pi^+\pi^-$, while the $D^*$ was identified by a slow pion.
We will now show that the $D^*$ chirality can be determined from a suitable angular decay 
distribution \cite{GPW}. We will only assume an $S$-wave $\rho^0\pi^-$ structure, 
without using the $a_1$ resonance shape and width, which involve a large uncertainty 
\cite{PDG}. A small $D$-wave correction can also be incorporated in the calculation.   

The decay amplitude for this process is written in terms of weak helicity amplitudes 
$H'_i$, in analogy with (\ref{A}),
\beq
A(\bar B^0\to D^{*+} \pi^-(p_1)\pi^-(p_2)\pi^+(p_3)) = \sum_{i=0,+,-} H'_i A_i~~.
\eeq
The strong amplitude $A_i$ involves two terms, corresponding to two possible ways 
of forming a $\rho$ meson from $\pi^+\pi^-$ pairs, each of which can be written 
in terms of two invariant amplitudes:
\beq\label{a1rhopi}
A(a_1(p,\varepsilon)\to \rho(p',\varepsilon') \pi) =
A(\varepsilon\cdot \varepsilon'^*) + B(\varepsilon\cdot p')
(\varepsilon'^*\cdot p)~,
\eeq
convoluted with the amplitude for $\rho^0(\varepsilon') \to \pi^+(p_i)\pi^-(p_j)$, which
is proportional to $\varepsilon'\cdot(p_i - p_j)$.
One finds  
\bea\label{a1amp}
& & A(a_1^-(p,\varepsilon) \to \pi^-(p_1) \pi^-(p_2) \pi^+(p_3)) \propto 
C(s_{13}, s_{23}) (\varepsilon\cdot p_1) +(p_1 \leftrightarrow p_2)~,\\
\label{C}
& & C(s_{13}, s_{23}) = [A + B m_{a_1} (E_3-E_2)] B_\rho (s_{23}) + 2A B_\rho (s_{13})~~,
\eea
where $s_{ij} = (p_i+p_j)^2,~B_{\rho}(s_{ij}) = (s_{ij} - m^2_{\rho} - im_{\rho}
\Gamma_{\rho})^{-1}$, and pion energies are given in the $a_1$ rest frame. The amplitudes 
$A$ and $B$ are related to $S$- and $D$-wave $\rho\pi$ amplitudes. 
When neglecting the small $D$-wave amplitude \cite{PDG}, they obey \cite{SD}
\beq
B = -A \left(1-\frac{m_\rho}{E_\rho}\right)\frac{E_\rho}{m_\rho \vec p_\rho\,^2}~~.
\eeq

Defining an angle $\theta$ between the normal to the $a_1$ decay plane and the 
direction opposite to the $D^*$ in the $a_1$ rest frame, one calculates the 
$B\to D^* 3\pi$ decay distribution,
\bea\label{dist}
\frac{d\Gamma}{ds_{13}ds_{23}d\cos\theta} & \propto &
|H'_0|^2 \sin^2\theta |\vec J|^2 + 
(|H'_+|^2 + |H'_-|^2)\frac12 (1 + \cos^2 \theta) |\vec J|^2\nonumber \\
& + & (|H'_+|^2 - |H'_-|^2) \cos\theta\,\mbox{Im}[(\vec J\times \vec J^*)\cdot \hat n]~~,
\eea
where
\beq\label{J}
\vec J = C(s_{13}, s_{23}) \vec p_1 + C(s_{23}, s_{13}) \vec p_2~~.
\eeq
A fit to the angular decay distribution enables separate measurements 
of the three terms $|H'_0|^2,~|H'_+|^2 + |H'_-|^2$ and $|H'_+|^2 - |H'_-|^2$. In 
particular, one can measure the P-odd up--down asymmetry of the $D^*$ momentum direction 
with respect to the $a_1$ decay plane,
\beq\label{As}
{\cal A} = \frac{3}{4} \frac{\langle \mbox{Im }[\hat n\cdot (\vec J\times \vec J^*)]
\mbox{sgn }(s_{13}-s_{23})\rangle}{\langle |\vec J|^2\rangle}
\frac{|H'_+|^2 - |H'_-|^2}{|H'_0|^2+|H'_+|^2 + |H'_-|^2}~~,
\eeq
which determines $|H'_+|^2 - |H'_-|^2$. Integration over the entire Dalitz plot gives
\beq
{\cal A} = -0.237 \frac{|H'_+|^2 - |H'_-|^2}{|H'_0|^2+|H'_+|^2 + |H'_-|^2}~~.
\eeq

In the heavy quark symmetry and factorization approximation \cite{factor}, using 
(\ref{H}) where $y \equiv (m_B^2+m_{D^*}^2-m_{a_1}^2)/2m_B m_{D^*} = 1.432,~
\epsilon \equiv m_{a_1}/(m_B - m_{D^*}) = 0.376$ , the results are
\beq
|H'_0| = 0.866~,~~~~ |H'_+| = 0.188~,~~~~|H'_-| = 0.463~.
\eeq
These values, which depend somewhat on $m_{a_1}$ and on neglecting corrections to 
factorization in $H'_{\pm}$, imply ${\cal A} = 0.042$. 
The sign of the asymmetry, which is not expected 
to change under these corrections, provides an unambiguous signature for a $V-A$ 
coupling in contrast to $V+A$. In the $a^-_1$ rest frame {\em the $\bar B^0$ and $D^{*+}$ 
prefer to move in the hemisphere opposite to $\vec{p}(\pi^-)_{\rm slow} \times 
\vec{p}(\pi^-)_{\rm fast}$}. Present statistics seem to be sufficient for determining this sign.
 
\section{Determining $2\beta + \gamma$ in CP asymmetries}
Both $\bar B^0 \to D^{*+}\rho^-$ and $\bar B^0 \to D^{*+}a^-_1$ belong to a class 
of processes, which also contains $\bar B^0 \to D^+\pi^-,~D^{*+}\pi^-,~D^+\rho^-$ \cite{ID}, 
from which the weak phase $2\beta + \gamma$ can be determined with no hadronic 
uncertainty. Using the well-measured 
value of $\beta$ \cite{psiKS} this would fix $\gamma$. The difficulty in these methods 
lies in having to measure  a very small time-dependent  
interference between $b\to c \bar u d$ and doubly-CKM-suppressed $\bar b \to \bar u 
c \bar d$ transitions, where $|V^*_{ub}V_{cd}/V_{cb}V^*_{ud}| \simeq 0.02$. 
In decays to $D^+\pi^-,~D^{*+}\pi^-,~D^+\rho^-$ the resulting 
analyses are sensitive to the doubly-CKM-suppressed rate, a precise measurement 
of which is extremely challenging. 
In the case of decays to two vector mesons, $\bar B^0 \to D^{*+}\rho^-$, one avoids the need to 
determine this small rate by using an interference between helicity amplitudes of CKM-allowed
and doubly-CKM-suppressed decays \cite{LSS}. This 
requires a detailed angular analysis in addition to time-dependent measurements. The 
feasibility of using an angular analysis for measuring the helicity amplitudes in the dominant 
CKM-allowed channel was demonstrated in \cite{CLEO}. It will be more difficult, but not 
impossible, to measure the time-dependent interference of helicity amplitudes with such disparate 
magnitudes. Here we will describe this method for determining $2\beta + \gamma$, first in
$\bar B^0 \to D^{*+}\rho^-$ \cite{LSS}, and then in $\bar B^0 \to D^{*+}a^-_1$ \cite{GPW}
where a discrete ambiguity in the weak phase can be resolved. Our considerations will not 
depend on an assumption of factorization.

\subsection{$\bar B^0(t) \to D^{*+}\rho^-$}
It is convenient to write the amplitude $A \equiv A(\bar B^0 \to D^{*+}(\to D^0\pi^+)
\rho^-(\to\pi^-\pi^0))$ in a linear polarization basis \cite{DQSTL}, in which the 
$D^*$ and $\rho$ transverse polarizations are either parallel or perpendicular to one another,
$H_{\parallel,\perp} = (H_+ \pm H_-)/\sqrt{2}$, and to similarly expand 
$a \equiv A(B^0 \to D^{*+}\rho^-)$ in terms of $h_{0,\parallel,\perp}$:
\bea\label{Aa}
A & = & \frac{3}{2\sqrt {2\pi}}(H_0g_0 + H_\parallel g_\parallel + iH_\perp g_\perp)~,~~
a = \frac{3}{2\sqrt {2\pi}}(h_0g_0 + h_\parallel g_\parallel + ih_\perp g_\perp)~,\\
g_0 & = & \cos\theta_1\cos\theta_2~,~g_\parallel = 
\frac{1}{\sqrt 2}\sin\theta_1\sin\theta_2\cos\phi~,~g_\perp = 
\frac{1}{\sqrt 2}\sin\theta_1\sin\theta_2\sin\phi~~.
\eea
The transversity amplitudes can be written as,
\beq
H_t = |H_t|\exp(i\Delta_t)~~,~~~~~~h_t = |h_t|\exp(i\delta_t)\exp(i\gamma)~~.
\eeq 

The time-dependent rate for $\bar B^0(t) \to D^{*+}\rho^-$ has the general form
\bea\label{Bt}
\Gamma(t) & \propto & e^{-\Gamma t}\left [(|A|^2 + |a|^2) + (|A|^2 - |a|^2)\cos\Delta mt\right .
+  \left. 2 {\rm Im}\left( e^{-2i\beta} A a^*\right)\sin \Delta mt\right ]\nonumber\\
& = & e^{-\Gamma t}\sum_{t \le t'}\left (\Lambda_{tt'} + \Sigma_{tt'}\cos\Delta mt 
+ \rho_{tt'}\sin\Delta mt \right )g_tg_t'~.
\eea 
Each of the coefficients in the sum can be measured by performing a time-dependent 
angular analysis. Denoting by $\Phi \equiv 2\beta + \gamma$, this determines the following quantities:
\bea\label{terms}
& & |H_t|^2~,~~~|H_0||H_\perp|\sin(\Delta_0 - \Delta_\perp)~,~~~|H_\parallel||H_\perp|\sin(\Delta_\parallel - 
\Delta_\perp)~~,\nonumber \\
& & |H_t||h_t|\sin(\Phi +\Delta_t -\delta_t )~,~~~~~~~~~~~~~~~~~~~~~~~~~~~~~(t = 0,\parallel,\perp )~~,
\nonumber \\
& & |H_\perp|h_0|\cos(\Phi  + \Delta_\perp - \delta_0) - 
|H_0|h_\perp|\cos(\Phi  + \Delta_0 - \delta_\perp)~~,\nonumber \\
& & |H_\perp|h_\parallel|\cos(\Phi  + \Delta_\perp - \delta_\parallel) - 
|H_\parallel|h_\perp|\cos(\Phi  + \Delta_\parallel - \delta_\perp)~~.
\eea
One does not rely on knowledge of the small $|h_t|^2$ terms \cite{LSS}, in which uncertainties would 
be large. Decays into the charge-conjugate state $D^{*-}\rho^+$ 
determine similar quantities, where $\Phi$ is replaced by $-\Phi$. It is then easy to show that 
this overall information is sufficient for determining $\sin\Phi$ {\em up to a sign ambiguity}.

\subsection{What is new in $\bar B^0(t) \to D^{*+}a^-_1$ ?}
The amplitudes $A'\equiv A(\bar B^0 \to D^{*+}(3\pi)^-_{a_1})$ and $a' \equiv A(B^0 \to 
D^{*+}(3\pi)^-_{a_1})$ are written in analogy with (\ref{Aa}): 
\bea
A' = \sum_{t=0,\parallel,\perp} H'_t A_t~~,~~~~
a' = \sum_{t=0,\parallel,\perp} h'_t A_t~~.
\eea
Instead of {\em real} functions $g_t$ of the angular variables, one has calculable {\em complex} 
amplitudes $A_t$ defined in Eq.~(\ref{a1amp}), which are functions of corresponding angles 
\cite{GPW}. One measures $\Gamma(\bar B^0(t) \to D^{*+}(3\pi)^-_{a_1})$ and   $\Gamma(\bar 
B^0(t) \to D^{*-}(3\pi)^+_{a_1})$ as a function of $\theta$ and an angle $\psi$ that defines the 
direction of the $D^*$ decay plane. The complex $A_t$, in contrast to the real $g_t$, imply 
that one can measure also interference terms between helicity amplitudes $H'_t$ and $h'_{t'}$ 
in which the cosines and sines in (\ref{terms}) are interchanged. 
This additional information has the effect of resolving the ambiguity in the sign of 
$\sin\Phi$ \cite{GPW}. 

The advantage of $B\to D^*a_1$ in determining {\em unambiguously} the CP-violating phase 
$2\beta + \gamma$ can be traced back to the parity-odd measurables that occur 
in this process but not in $B \to D^*\rho$. As noted, $|H'_+|^2 - H'_-|^2 = 2{\rm Re}
(H'_\parallel H'^*_\perp)$ is P-odd, and so is ${\rm Im}[e^{2i\beta}(H'_\parallel h'^*_\perp + 
H_\perp h^*_\parallel )]$. These terms, which do not occur in the time-dependent rate of 
$\bar B^0 \to D^{*+}\rho^-$, do occur in $\bar B^0(t) \to D^{*+}a_1^-$ multiplying a P-odd function 
of $\theta$, $\cos\theta\,\mbox{Im}[(\vec J\times \vec J^*)\cdot \hat n]$. A practical advantage of 
$\bar B^-\to D^{*+}a^-_1$ over $\bar B^0 \to D^{*+}\rho^-$ is the occurrence of only charged pions
in the first process. A slight disadvantage of the first process may be an intrinsic uncertainty 
in the amplitudes $A_t$ obtained in (\ref{a1amp}). 

\section{Conclusion}
Parity-odd measurables in hadronic and photonic $B$ decays were shown to test the chiral 
structure of weak $b$ quark couplings at tree level and at the one-loop level, respectively. 
Time-dependent CP asymmetries in $B \to D^* a_1$ complement measurements in $B \to D^*\rho$, and 
resolve a discrete ambiguity in a clean determination of the CP-violating phase $2\beta + \gamma$.

\medskip
I am grateful to the CERN Theory Division for its kind hospitality.

\def \ajp#1#2#3{Am.\ J. Phys.\ {\bf#1}, #2 (#3)}
\def \apny#1#2#3{Ann.\ Phys.\ (N.Y.) {\bf#1}, #2 (#3)}
\def \app#1#2#3{Acta Phys.\ Polon. {\bf#1}, #2 (#3)}
\def \arnps#1#2#3{Ann.\ Rev.\ Nucl.\ Part.\ Sci.\ {\bf#1}, #2 (#3)}
\def \art{and references therein}
\def \cmts#1#2#3{Comments on Nucl.\ Part.\ Phys.\ {\bf#1}, #2 (#3)}
\def \cn{Collaboration}
\def \ib{{\it ibid.}~}
\def \ibj#1#2#3{~{\bf#1}, #2 (#3)}
\def \ijmpa#1#2#3{Int.\ J.\ Mod.\ Phys.\ A {\bf#1}, #2 (#3)}
\def \ite{{\it et al.}}
\def \jhep#1#2#3{JHEP {\bf#1}, #2 (#3)}
\def \jpb#1#2#3{J.\ Phys.\ B {\bf#1}, #2 (#3)}
\def \kdvs#1#2#3{{Kong.\ Danske Vid.\ Selsk., Matt-fys.\ Medd.} {\bf #1}, No.\
#2 (#3)}
\def \mpla#1#2#3{Mod.\ Phys.\ Lett.\ A {\bf#1}, #2 (#3)}
\def \nat#1#2#3{Nature {\bf#1}, #2 (#3)}
\def \nc#1#2#3{Nuovo Cim.\ {\bf#1}, #2 (#3)}
\def \nima#1#2#3{Nucl.\ Instr.\ Meth.\ A {\bf#1}, #2 (#3)}
\def \npb#1#2#3{Nucl.\ Phys.\ B~{\bf#1}, #2 (#3)}
\def \npps#1#2#3{Nucl.\ Phys.\ Proc.\ Suppl.\ {\bf#1}, #2 (#3)}
\def \os{XXX International Conference on High Energy Physics, Osaka, Japan,
July 27 -- August 2, 2000}
\def \PDG{Particle Data Group, D. E. Groom \ite, \epjc{15}{1}{2000}}
\def \pisma#1#2#3#4{Pis'ma Zh.\ Eksp.\ Teor.\ Fiz.\ {\bf#1}, #2 (#3) [JETP
Lett.\ {\bf#1}, #4 (#3)]}
\def \pl#1#2#3{Phys.\ Lett.\ {\bf#1}, #2 (#3)}
\def \pla#1#2#3{Phys.\ Lett.\ A {\bf#1}, #2 (#3)}
\def \plb#1#2#3{Phys.\ Lett.\ B {\bf#1}, #2 (#3)}
\def \prl#1#2#3{Phys.\ Rev.\ Lett.\ {\bf#1}, #2 (#3)}
\def \prd#1#2#3{Phys.\ Rev.\ D\ {\bf#1}, #2 (#3)}
\def \prp#1#2#3{Phys.\ Rep.\ {\bf#1}, #2 (#3)}
\def \ptp#1#2#3{Prog.\ Theor.\ Phys.\ {\bf#1}, #2 (#3)}
\def \rmp#1#2#3{Rev.\ Mod.\ Phys.\ {\bf#1}, #2 (#3)}
\def \rp#1{~~~~~\ldots\ldots{\rm rp~}{#1}~~~~~}
\def \yaf#1#2#3#4{Yad.\ Fiz.\ {\bf#1}, #2 (#3) [Sov.\ J.\ Nucl.\ Phys.\
{\bf #1}, #4 (#3)]}
\def \zhetf#1#2#3#4#5#6{Zh.\ Eksp.\ Teor.\ Fiz.\ {\bf #1}, #2 (#3) [Sov.\
Phys.\ - JETP {\bf #4}, #5 (#6)]}
\def \zpc#1#2#3{Z.\ Phys.\ C {\bf#1}, #2 (#3)}
\def \zpd#1#2#3{Z.\ Phys.\ D {\bf#1}, #2 (#3)}


\begin{thebibliography}{99}
\bibitem{Cab} N. Cabibbo, \prl{10}{531}{1963}
.
\bibitem{KM} M. Kobayashi and T. Maskawa, \ptp{49}{652}{1973}.

\bibitem{sanda} A. B. Carter and A. I. Sanda, \prd{23}{1567}{1981}; I. I. Bigi 
and A. I. Sanda, \npb{193}{85}{1981}.

\bibitem{gold} M. Gronau, \prl{63}{1451}{1989}; D. London and R. L. Peccei, 
\plb{223}{257}{1989}.

\bibitem{psiKS} BaBar \cn, B. Aubert \ite, \prl{87}{091801}{2001}; 
\prd{66}{032003}{2002}; \prl{89}{201802}{2002}; Belle \cn, K. Abe \ite, 
\prl{87}{091802}{2001}; \prd{66}{032007}{2002}; \prd{66}{071102}{2002}.
  
\bibitem{pipi} BaBar \cn, B. Aubert \ite, \prl{89}{281802}{2002}; Belle \cn, 
K. Abe \ite, hep-ex/0301032, submitted to Phys. Rev. D.

\bibitem{DK} Belle \cn, K. Abe \ite, hep-ph/0207012,
submitted to Phys. Rev. Lett.; BaBar \cn, B. Aubert \ite, hep-ph/0207087.

\bibitem{MGR} M. Gronau and J. L. Rosner, \prd{65}{013004}{2001}; 
\prd{65}{093012}{2002}; \prd{66}{053003}{2002}; M. Gronau, hep-ph/0211282, 
\plb{557}{198}{2003}.

\bibitem{Bs} R. Fleischer, \plb{459}{306}{1999}; M. Gronau and J. L. Rosner, 
\plb{482}{71}{2000}; \prd{65}{113008}{2002}.

\bibitem{isospin} M. Gronau and D. London, \prl{65}{3381}{1990}; 
H. J. Lipkin, Y. Nir, H. R. Quinn and A. Snyder, \prd{44}{1454}{1991}.
M. Gronau, \plb{265}{389}{1991}.  

\bibitem{GHLR} M. Gronau, O. F. Hern\'andez, D. London, and J. L. Rosner,
\prd{50}{4529}{1994}; \prd{52}{6374}{1995}.

\bibitem{BBNS} M. Beneke, G. Buchalla, M. Neubert, and C. T. Sachrajda, 
\prl{83}{1914}{1999}; \npb{606}{245}{2001};
Y. Y. Keum, H. N. Li and A. I. Sanda, \plb{504}{6}{2001}; \prd{63}{054008}{2001}. 

\bibitem{LR} J. C. Pati and A. Salam, \prd{10}{275}{1974}; R. N. Mohapatra 
and J. C. Pati, \prd{11}{566}{1975}.

\bibitem{PDG} Particle Data Group, K. Hagiwara {\em et al.}, \prd{66}{010001}{2002}.

\bibitem{RMAP} R. N. Mohapatra and A. P\'erez-Lorenzana, \prd{66}{135005}{2002}. 

\bibitem{MG} M. Gronau, B Decays, ed. S. Stone (World Scientific, Singapore, 1994), 
p. 644.

\bibitem{CP} M. Gronau, Proc. Fourth Int. Conf. on B Physics and CP 
Violation, Ago Town, Japan, Feb. 2001, ed. T. Ohshima and A. I. Sanda (World 
Scientific 2001),~p. 13;~R. Fleischer, Proc. Ninth Int. Sym. on Heavy Flavor 
Physics, Pasadena, California, 2001, AIP Conf. Proc.~{\bf 618},~266~(2002); 
M. Neubert, Proc. 20th Int. Sym. on Lepton and Photon Interactions 
at High Energies, Rome, Italy, 2001, \ijmpa{17}{2936}{2002}.

\bibitem{hurth} T. Hurth, hep-ph/0212304, CERN-TH/2002-264.

\bibitem{GP} M. Gronau and D. Pirjol, \prd{66}{054008}{2002}.
 
\bibitem{LRM} K. Fujikawa and A. Yamada, \prd{49}{5890}{1994}; K. S. Babu, K.
Fujikawa and A. Yamada, \plb{333}{196}{1994}; P.~Cho and M.~Misiak, \prd{49}
{5894}{1994}.

\bibitem{SUSY} L. Everett, G. L. Kane, S. Rigolin, L. T. Wang and T. T. Wang,
\jhep{0201}{022}{2002}. 
 
 
\bibitem{AGS} D. Atwood, M. Gronau and A. Soni, \prl{79}{185}{1997}.

\bibitem{Kim} D. Melikhov, N. Nikitin and S. Simula, \plb{442}{381}{1998};
F. Kr\"uger, L. M. Sehgal, N. Sinha and R. Sinha, \prd{61}{114028}{2000};
C. S. Kim, Y. G. Kim, C. D. L\"u and T. Morozumi, \prd{62}{034013}{2000}.

\bibitem{GrPi} Y. Grossman and D. Pirjol, \jhep{0006}{029}{2000}.

\bibitem{MaRe} T. Mannel and S. Recksiegel, \app{B28}{2489}{1997}. 

\bibitem{HK} G. Hiller and A. Kagan, \prd{65}{074038}{2002}.

\bibitem{CLEOK**} CLEO \cn, T. E. Coan \ite, \prl{84}{5283}{2000}.

\bibitem{Belle} Belle Collaboration, S. Nishida \ite, \prl{89}{231801}{2002}.

\bibitem{GGPR} M. Gronau, Y. Grossman, D. Pirjol and A. Ryd,
\prl{88}{051802}{2002}.

\bibitem{charged} Decays to $K\pi\pi\gamma$ with only charged pions involve 
a larger theoretical uncertainty \cite{GP}.

\bibitem{BR} T. Altomari, \prd{37}{677}{1988}; A. Ali, T. Ohl and 
T. Mannel, \plb{298}{195}{1993}; D. Atwood and A. Soni, \zpc{64}{241}{1994}; 
S. Veseli and M. G. Olsson, \plb{367}{309}{1996}; 
\plb{495}{309}{2000};
D. Ebert, R. N. Faustov, V. O. Galkin and H. Toki, \prd{64}{054001}{2001}.

\bibitem{CLEOV-A} CLEO \cn, S. Sanghera \ite, \prd{47}{791}{1993}; CLEO \cn, J. E. 
Duboscq \ite, \prl{76}{3898}{1996}.

\bibitem{GW2} M. Gronau and S. Wakaizumi, \plb{280}{79}{1992}.

\bibitem{KG} J. G. K\"orner and G. R. Goldstein, \plb{89}{105}{1979}.

\bibitem{CLEO} CLEO \cn, S. E. Csorna \ite, Cornell Report CLNS 03/1813, 
hep-ex/0301028; CLEO \cn, M. S. Alam \ite, \prd{50}{43}{1994}.

\bibitem{BBNS2} M. Beneke, G. Buchalla, M. Neubert and C. T. Sachrajda, 
\npb{591}{313}{2000}; C. W. Bauer, D. Pirjol and I. W. Stewart, 
\prl{87}{201806}{2001}.

\bibitem{factor} J. L. Rosner, \prd{42}{3732}{1990}.

\bibitem{BABARa1} BABAR \cn, B. Aubert \ite, Conference Report BABAR-CONF-02/10, 
hep-ex/0207085; ARGUS \cn, H. Albrecht \ite, \zpc{48}{543}{1990}; 
CLEO \cn, M. S. Alam \ite, \prd{45}{21}{1992}, {\bf 50}, 43 (1994).

\bibitem{GPW} M. Gronau, D. Pirjol and D. Wyler, \prl{90}{051801}{2003}.

\bibitem{SD} N. Isgur, C. Morningstar and C. Reader, \prd{39} {1357}{1989}; 
M. Feindt, \zpc{48}{681}{1990}.

\bibitem{ID} I. Dunietz, \plb{427}{179}{1998}; D. A. Suprun, C. W. Chiang 
and J. L. Rosner, \prd{65}{054025}{2002}. 


\bibitem{LSS} D. London, N. Sinha and R. Sinha, \prl{85}{1807}{2000}.

\bibitem{DQSTL} I. Dunietz, H. R. Quinn, A. Snyder, W. Toki and H. J. Lipkin, 
\prd{43}{2193}{1991}.

\end{thebibliography}
\end{document}